\documentclass[10pt,conference]{IEEEtran}

\usepackage{amssymb}
\usepackage{graphicx}
\usepackage{pifont}

\begin{document}
\linespread{0.92}

\title{Evaluating the Indistinguishability of Logic Locking using K-Cut Enumeration and Boolean Matching
}

\author{\IEEEauthorblockN{Jonathan Cruz and Jason Hamlet}
\IEEEauthorblockA{Sandia National Laboratories, NM, USA\\
Email: jwcruz@sandia.gov}\\
}

\maketitle

\begin{abstract}
Logic locking as a solution for semiconductor intellectual property (IP) confidentiality has received considerable attention in academia, but has yet to produce a viable solution to protect against known threats. In part due to a lack of rigor, logic locking defenses have been historically short-lived, which is an unacceptable risk for hardware-based security solutions for critical systems that may be fielded for decades. Researchers have worked to map the concept of cryptographic indistinguishability to logic locking, as indistinguishability provides strong security guarantees.  In an effort to bridge theory and practice, we highlight recent efforts that can be used to analyze the indistinguishability of logic locking techniques, and propose a new method of evaluation based on comparing distributions of $k$-cuts, which is akin to comparing against a library of sub-functions. We evaluate our approach on several different classes of logic locking and show up to 92\% average accuracy in correctly identifying which design was locked, even in the presence of resynthesis, suggesting that the evaluated locks do not provide indistinguishability.
\end{abstract}

\section{Introduction}

The need for secure countermeasures against threats to semiconductor Intellectual Property (IP) confidentiality and integrity during manufacturing has lead to several active research areas in countermeasure development, with logic locking a primary thrust. Logic locking introduces key controlled circuitry that modifies circuit functionality or removes functionality all together until the correct unlocking key is provided~\cite{kamali2022advances}. In addition to obscuring circuit functionality and preventing adversarial key recovery, logic locking techniques also seek to defend against attacks targeted at recovering the original circuit design. 

Logic locking approaches have undergone short proposal-to-defeat cycles.  This calls into question their practicality, especially given that the locks are implemented in hardware and cannot be updated after deployment. Moreover, much of the current security analysis has been narrowly focused on a subset of the threat space, with most locking approaches only considering oracle-guided or SAT attack resilience. Researchers have worked towards establishing formalisms and security primitives to improve rigor, with indistinguishable logic locking a common theme~\cite{beerel2022towards,el2022locked}. Notionally, indistinguishability provides strong security guarantees, which makes this an attractive property for ensuring the confidentiality of a locked circuit. However, there remains uncertainty about how to evaluate the indistinguishability of a lock.

Recent locking analysis approaches evaluate indistinguishability directly by comparing a locked design to another locked design or a potentially similar unlocked design~\cite{hamlet2025perspective,dasgupta2025latte,akib2025knowledge}. However, these approaches are either heavily dependent on the underlying structure for the comparison or are unnecessarily self-restricting, ignoring portions of the design affected by the locking functions. To resolve these shortcomings, we propose a new method to evaluate the indistinguishability of logic locks by constructing a corpus of k-input logic cones, or $k$-cuts, in a design to use for comparison to known designs. We employ NPN Boolean equivalence class matching to account for potential perturbations introduced to the matching process by the logic locks, which can change both the structure and function of the original design. By comparing the NPN equivalence of cut distributions, our approach considers both of these potential changes.
  
The main contributions of this manuscript are as follows:
\begin{itemize}
    \item We present a novel $k$-cut enumeration approach for evaluating indistinguishability modeled after the chosen plaintext attack.
    \item We perform comprehensive experimentation on several state of the art locking techniques.
\end{itemize}
The rest of the paper is organized as follows: Section~\ref{sec:background} provides a brief introduction to logic locking and indistinguishability; Section~\ref{sec:related_works} discusses related works for evaluating indistinguishability; Section~\ref{sec:threat_model} introduces our threat model;  Section~\ref{sec:method} describes our methodology; Section~\ref{sec:evaluation} discusses our experimental results; Section~\ref{sec:discussion} highlights insights and limitations of the proposed approach and finally Section~\ref{sec:conclusion} concludes the paper.

\section{Background}
\label{sec:background}

\subsection{Logic Locking}
Logic locking techniques can be categorized by the portion of the design they are applied to and their fundamental methodologies. Combinational logic locking techniques apply to purely combinational circuits as well as the combinational portions of sequential circuits. The combinational gates are locked with key structures such that the correct function is prevented unless the correct key is applied. Techniques that add locking structures or additional, incorrect combinational paths include bitwise, point-function (sometimes called corrupt and correct), cyclic, and routing approaches~\cite{kamali2022advances}. There are also redaction based locking techniques that remove design logic in place of programmable logic~\cite{kamali2022advances}. Sequential logic locking approaches typically either lock the sequential elements, such as finite state machines (FSMs), in a circuit by adding transitions from them to additional locking states, or introduce new state machines that must be traversed in a specified way to enter the functional FSM~\cite{kamali2022advances}. The focus of this work is on evaluating the indistinguishability of combinational logic locks.

\subsection{Indistinguishability}
Several papers have introduced formalized security models for logic locking in an attempt to add more rigor to the field.  A promising idea has been mapping cryptographic indistinguishability to logic locking to allow reasoning about a lock's security~\cite{beerel2022towards,el2022locked}. In cryptography, establishing the indistinguishability of a cipher is often expressed on the output of a security game. While several variations of this game exist, our focus is on indistinguishability under chosen plaintext attacks for symmetric ciphers. The security game is described as follows: the defender possesses an encryption oracle with a secret key unknown to the attacker. The attacker may perform some bounded number of encryptions using this oracle. Eventually, the attacker submits two plaintexts to the defender. The defender chooses one plaintext at random to encrypt, and returns the ciphertext to the attacker, who must now determine which plaintext was encrypted. If the attacker succeeds with only some negligible advantage over random guessing, then the cryptosystem is indistinguisable under chosen plaintext attacks.

With logic locking, there are two forms of plaintext, the original design and the correct functional input-to-output (I/O) pairs. Access to the original design implies access to the correct functional I/O pairs. In theory, an indistinguishable lock would produce a circuit that is indistinguishable from any other circuit under any level of analysis, either structural or functional. Hamlet et al., built off these formalisms of locking indistinguishability and outlined necessary conditions for secure locking~\cite{hamlet2025perspective}. The authors specify that a lock should be indistinguishable under structural and functional analysis, including sub-components, within some nominal error. In this work, we propose a robust way to evaluate indistinguishability through both structural and functional analysis.

\section{Related Works}
\label{sec:related_works}
\begin{figure} [!t]
    \centering
    \includegraphics[width=0.99\columnwidth]{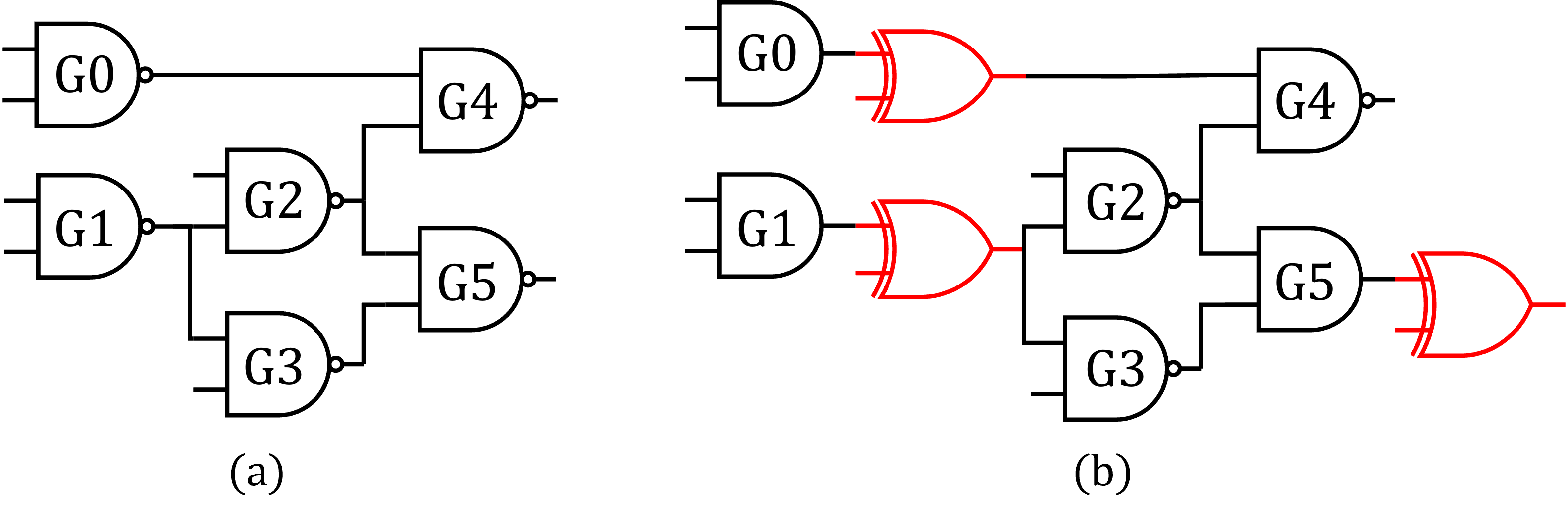}
    \caption{Example highlighting key gate (red) placement and potential structural and functional transformations from locking. (a) The original c17 benchmark with 3-cut distribution ((G5,G3), (G5,G2), (G5,G3,G2), (G3,G1), (G2,G1), (G4,G2), (G4,G0)). (b) c17 locked with TRLL which transforms the original design by replacing inverters with XOR key gates with key value 1. The 3-cut distribution excluding key gates is ((G5,G3), (G5,G2), (G5,G3,G2), (G4,G2)).}
    \vspace{-0.2in}
    \label{fig:locking_example}
\end{figure}

Recently, a new class of oracle-less attacks that use databases of known designs have emerged as a defeat to combinational locking techniques~\cite{hamlet2025perspective,dasgupta2025latte,akib2025knowledge}. Fundamentally, the approaches highlight the lack of structural obfuscation introduced by locking techniques. In~\cite{hamlet2025perspective}, the authors propose a structural analysis approach comparing the distribution of graph edit distances (GED) between locked and unlocked designs. More importantly, the paper introduced k*-security to quantify the amount of unperturbed logic from the original design that exists in the locked design. LATTE~\cite{dasgupta2025latte} is an approach that uses a database of known designs and compares to the locked design under test (L-DUT) using structural analysis based on a similarity metric (s-metric) comprised of number of PIs, number of vertices, logic depth, and edge connectivity. The LATTE attack requires a transformation tool to generate multiple variants of the designs within the library to compare with the L-DUT highlighting the dependence of the s-metric on the structures available in the library.  Moreover, it does not differentiate the locking structures from the original design, further complicating comparison. KOLA~\cite{akib2025knowledge} is a technique that compares the unlocked portions of the L-DUT to a database of unlocked designs using weak Boolean division. However, this approach to comparing unaffected or ``exposed" portions of the design is hampered by high overhead locks, or locks simply placed towards the primary inputs of the design. We argue this suggested mitigation is insufficient and still leaves the fundamental problem unaddressed -- the original design or portions of the design that contribute to its identity, are still mostly intact. To further highlight the issue, consider the example shown in Figure~\ref{fig:locking_example}. In this example the c17 benchmark was locked with TRLL, which transformed G0, G1 and G5 NAND gates into AND gates. The KOLA flow does not consider gates in the fanout of locking gates, therefore, only gates G0 and G1 remain for analysis. Because these gates have been changed by the lock, they will not match the original design. However, of the four gates in the fanout of the locking gates, three are unchanged and can still be used for analysis. 
A summary comparing our proposed approach to other related works is shown in Table~\ref{table:related_works}
\begin{table}[t] 
	\centering
	\noindent
    \caption{Summary of comparison between our proposed approach and state of the art}
	\centerline{\resizebox{1\columnwidth}{!}{
	\begin{tabular}{llll}
		\hline
		  Approach   & Isolate Lock &Comp. Level & Comp. Type  \\
		 \hline
		GED KS-test~\cite{hamlet2025perspective} & \ding{55} & Cut & Structural (GED) \\ 
        LATTE~\cite{dasgupta2025latte} &  \ding{55}& Cut & Structural (S-Metric) \\ 
        KOLA~\cite{akib2025knowledge} & \checkmark & Unobf. Logic & Functional (Weak Division)\\ & & Cone & \\ 
        K-Cut Enum. & \checkmark &Cut & Functional (NPN)\\  (Proposed) & & & Structural (Cut Enum) \\ 
        
		\hline
		\hline
	\end{tabular}}}
	
 \label{table:related_works}
 \vspace{-0.2in}
\end{table}

\begin{figure}[t]
        \includegraphics[width=0.99\columnwidth]{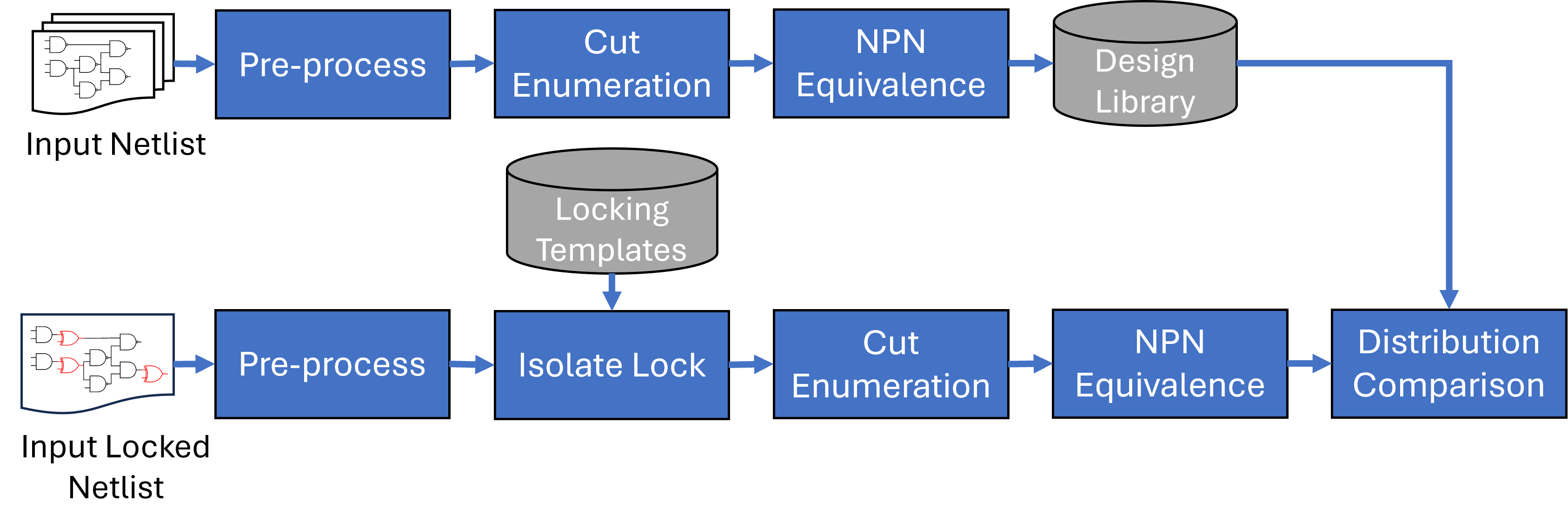}
        \caption{Overall flow for $K$-Cut Enumeration Chosen Design Plaintext Analysis.}
        \label{fig:flow}
        \vspace{-0.15in}
\end{figure}

\section{Threat Model}
\label{sec:threat_model}
Our threat model focuses on a knowledgeable oracle-less adversary attempting to reverse engineer the design. The adversary has knowledge of the locking method employed by the defender and a corpus of different circuits and classes of designs. The adversary's goal is to match the locked circuit to a known unlocked circuit.

\section{Indistinguishability under Chosen Design Plaintext Attack} 
\label{sec:method}
As we mentioned above, our approach to using cut distribution analysis to assess indistinguishability is modeled after the chosen plaintext attack in cryptography, but applied to logic locking. In our case, the chosen plaintexts are unlocked designs.

Given a locked design under test (L-DUT) as input, the major steps of the proposed analysis flow are as follows: 1) pre-process L-DUT, 2) isolate the logic directly associated with the lock from the logic of the original design, 3) for some k, enumerate $k$-cuts for every non-locking gate in the design, 4) construct a database of $k$-cut equivalence classes and finally, 5) compare the database of cut equivalence classes to the corpus of $k$-cut enumerations extracted from known designs.  
The main flow is illustrated in Figure~\ref{fig:flow}. We further describe each of the major steps below.

\subsection{Preliminaries}
\begin{itemize}
    \item A combinational netlist can be represented as a directed acyclic graph, $G$, with $V$ vertices as gates and $E$ edges as wires $(V,E)\in G$.
    \item A $cut$ is a collection of connected gates rooted at a gate $v\in V$.
    \item A $k$-$cut$ is a cut with at most $k$ unique inputs. 
\end{itemize}
\subsection{Design Pre-processing and Data Labeling}
A design can be deliberately unoptimized pre-locking~\cite{alrahis2021unsail} or mapped to any number of technology libraries post-locking to add to the structural diversity and ultimately hinder structural analysis. Therefore, the design pre-processing stage is an important step to normalize the input netlist before locking component identification and subsequent cut enumeration. We pre-process the L-DUT by resynthesizing the design using a minimal technology library consisting of 2-input gates. 

Next, we attempt to label gates that belong to the lock. As our threat model provides the adversary knowledge of the locking technique we use a database of functional templates to match against. The adversary can also use knowledge of the key inputs to further refine the matching process, which is a common assumption in literature~\cite{akib2025knowledge}. From a key input, the adversary can enumerate cuts and label them as part of the lock if they match against the locking template database. The remainder of the design not labeled as belonging to the lock is used during cut enumeration to create a design signature.

\subsection{Cut Enumeration}
\label{sec:cut_enum}
To enumerate cuts, we traverse the graph and identify all possible $k$-cuts for a given $k$ from every gate with the intuition that the collection of cuts or sub-functions in a design can be used as an approximation for establishing a design signature. When performing cut enumeration on the L-DUT, we stop exploring a path once an identified locking gate is reached.

The cut enumeration process is inherently dependent on the number of gates and their connections. With more gates and larger $k$ there are more possible cuts~\cite{mishchenko2007combinational}, which can lead to lower similarity scores when comparing two cut distributions. Therefore, when enumerating $k$-cuts for the purpose of constructing a design signature, we select only a subset of cuts as representative for the enumerated cut population from a source node. This selection is done by prioritizing cuts that contain more gates, excluding buffers and inverters. The expected distribution size for a given design is now $n\times |V|$, where $n$ is the subset chosen. The synthesis run during pre-processing also helps to reduce structural variations.

We then compute and store the NPN equivalence class for each cut selected. Two Boolean functions are NPN equivalent if they are equal under the negation of inputs, permutation of inputs, or negation of outputs. This database of NPN equivalence classes of $k$-cuts serves as the design signature. Given two NPN cut distributions, we compare their similarity using Jaccard set similarity.

We use NPN equivalence for two reasons: 1) to account for potential imprecision in cut enumeration (input order or grabbing inverters) and 2) to reconcile that we are, in effect, coarsely removing the locking structures during the labeling phase, which may introduce unwanted negations to the original design. For bitwise locking approaches, NPN matching is particularly effective since these locks introduce one bit flip per key gate. Due to this, depending on where the key gate is introduced, we can carve out cuts that capture the original design function. From Figure~\ref{fig:locking_example}, with 3-cut enumeration and NPN matching we can still correctly account for 4 out of 7 gates (G2, G3, G4, G5) between the locked design and the original. For more complicated corruption approaches such as the corrupt and correct techniques, the extent to which NPN matching can correctly identify the whole or portions of the functionally stripped circuit is dependent upon the hamming distance metric used during locking. 
\section{Evaluation}
\label{sec:evaluation}

\begin{table*}[th!] 
	\centering
	\noindent
    \caption{K-Cut Enumeration Accuracy}
	\centerline{\resizebox{0.95\textwidth}{!}{
	\begin{tabular}{ll|llll|llll|llll}
		\hline
		  Lock   & Key Size & \multicolumn{4}{c}{4-cut} & \multicolumn{4}{c}{6-cut} & \multicolumn{4}{c}{8-cut}    \\
               & & Top 1 & Top 5 & Top 10 & Top 20 & Top 1 & Top 5 & Top 10 & Top 20 & Top 1 & Top 5 & Top 10 & Top 20 \\
              
		 \hline
        TRLL & 32 &  0.875 & 1 & 1 & 1 & 1 & 1 & 1 & 1 & 1 & 1 & 1 & 1 \\
            & 64 &  1 & 1 & 1 & 1 & 1 & 1 & 1 & 1 & 1 & 1 & 1 & 1 \\
            & 128 & 0.75 & 0.75 & 0.75 & 0.75 & 0.75 & 0.75 & 0.75 & 0.75 & 0.75 & 0.875 & 0.875 & 0.75 \\
         SFLLHD & 32&           0.875 & 0.875 & 0.75 & 0.75 & 1 & 1 & 1 & 1 & 1 & 1 & 1 & 1 \\
                & 64 &    1 & 0.5 & 0.5 & 0.5 & 1 & 1 & 1 & 1 & 1 & 1 & 1 & 1 \\
                & 128 &    0.5 & 0.5 & 0.5 & 0.5 & 1 & 1 & 1 & 1 & 1 & 1 & 1 & 1 \\
          MUX & 32 &          0.875 & 1 & 1 & 1 & 1 & 1 & 1 & 1 & 1 & 1 & 1 & 1 \\
              & 64 &      0.75 & 0.875 & 0.875 & 0.875 & 1 & 0.875 & 1 & 0.875 & 0.75 & 0.75 & 0.875 & 0.875 \\
              & 128 &      0.5 & 0.667 & 0.667 & 0.667 & 0.667 & 0.5 & 0.5 & 0.667 & 0.667 & 0.5 & 0.5 & 0.667 \\
          LUT & 32 &          1 & 0.857 & 0.857 & 0.857 & 0.857 & 1 & 0.857 & 0.857 & 0.857 & 1 & 1 & 1 \\
               & 64 &     0.857 & 0.857 & 0.857 & 0.857 & 0.857 & 0.857 & 0.857 & 0.857 & 0.857 & 0.857 & 0.857 & 0.857 \\
               & 128 &     0.75 & 0.75 & 0.75 & 0.75 & 0.75 & 1 & 0.75 & 1 & 0.75 & 1 & 1 & 1 \\

        \hline 
         Average\textsuperscript{\textdagger} &   & 0.829 & 0.855 & 0.842 & 0.842 & 0.908 & 0.908  & 0.895 & 0.908 & 0.882 & 0.908 & 0.921 & 0.921 \\
		\hline
	\end{tabular}}}
	\raggedright{\textdagger Average accuracy across all locked test articles.}
 \label{table:main_results}
 \vspace{-0.2in}
\end{table*}

\begin{table}[] 
	\centering
	\noindent
    \caption{Summary of Locked Benchmark Overheads}
	\centerline{\resizebox{1\columnwidth}{!}{
	\begin{tabular}{l|lll|lll|lll|lll}
		\hline
		  Benchmark   &  \multicolumn{12}{c}{Locking Method}     \\
               & \multicolumn{3}{c}{TRLL}  & \multicolumn{3}{c}{SFLLHD}  & \multicolumn{3}{c}{MUX}  & \multicolumn{3}{c}{LUT} \\
               & 32 & 64 & 128 & 32 & 64 & 128 & 32 & 64 & 128 & 32 & 64 & 128 \\
		 \hline
		c432 & 1.168 & 1.363 & 2.197 & 1.101 &  &  & 1.593 & 2.185 &  & 1.269 & 1.560 & 2.217 \\
        c499 & 1.157 & 1.286 & 1.557 & 1.097 &  &  & 1.520 & 2.041 &  & 1.235 & 1.487 &  \\
        c880 & 1.086 & 1.187 & 1.378 & 1.081 &  &  & 1.313 & 1.626 & 2.252 & 1.140 & 1.285 & 1.594\\
        c1355 & 1.071 & 1.163 & 1.315 & 1.068 & &  & 1.217 & 1.434 & 1.868 & &  &  \\
        c1908 & 1.042 & 1.092 & 1.166 & 1.054 &  &  & 1.143 & 1.286 & 1.573 & 1.063 & 1.128 &  \\
        c2670 & 1.032 & 1.061 & 1.126 & 1.045 & 1.075 & 1.060 & 1.106 & 1.213 & 1.425 & 1.047 & 1.094 & 1.191 \\
        c3540 & 1.021 & 1.050 & 1.081 & 1.034 &  &  & 1.073 & 1.145 & 1.291 & 1.032 & 1.064 &  \\
        c5315 &  1.013 & 1.025 & 1.052 & 1.024 & 1.038 & 1.054 & 1.048 & 1.096 & 1.192 & 1.021 & 1.042 & 1.085 \\
       
		\hline
		\hline
	\end{tabular}}}
	
 \label{table:overheads}
 \vspace{-0.15in}
\end{table}

\subsection{Experimental Setup}

We apply the following locks on the original ISCAS'85 benchmark suite with variable key lengths (32, 64, 128) using the open-source framework circuitgraph~\cite{sweeney2020circuitgraph}: TRLL~\cite{limaye2020thwarting}, MUX Lock~\cite{rajendran2013fault}, LUT Lock~\cite{kamali2018lut}, and SFLL-HD~\cite{yasin2017provably}. For SFLL-HD we choose a Hamming Distance (HD) parameter of half of the key size, suggested by the original authors to appropriately tradeoff between structural and SAT resilience~\cite{yasin2017provably}. A summary of locked benchmarks and their area overheads are shown in Table~\ref{table:overheads}. The lack of a benchmark and key size combination implies the open-source tool was unable to generate the locked benchmark. 

After locking, the benchmarks are resynthesized in the pre-processing step described in Section~\ref{sec:method} to a simple standard cell library that consists of 2-input AND, NAND, OR, NOR, XOR, XNOR and 1-input BUF, INV gates. The unlocked ISCAS'85 benchmarks are resynthesized to the same standard cell library and used as the database of unlocked designs with which to compare the L-DUT. We remove c499 from the reference benchmarks because it is functionally equivalent to c1355. For cut enumeration, we compute distributions for 4-, 6-, and 8-cuts. We search up to 10000 cuts from a root gate and store the largest N (top-1, -5, -10, -20) cuts identified. The size of a cut is computed as the number of 2-input gates.
Our cut analysis technique is implemented in Python. We implement an approximate NPN Boolean matching heuristic using the technique described in~\cite{zhang2021enhanced}.  

\subsection{Results}

Table~\ref{table:main_results} presents the accuracy for correctly identifying the benchmark that was locked using 4-, 6-, and 8-cut enumeration. The highest average accuracy of 92\% was achieved with top-10 and top-20 cut selection for 8-cuts. From Table~\ref{table:main_results}, we observe a general trend where as the key size increases, the accuracy of $k$-cut enumeration decreases. Intuitively, more key gates dispersed through the design equates to more points of stoppage during our cut enumeration (as described in Section~\ref{sec:cut_enum}), disrupting the comparison.  

\begin{figure*}
        \centering
        \includegraphics[width=0.97\textwidth]{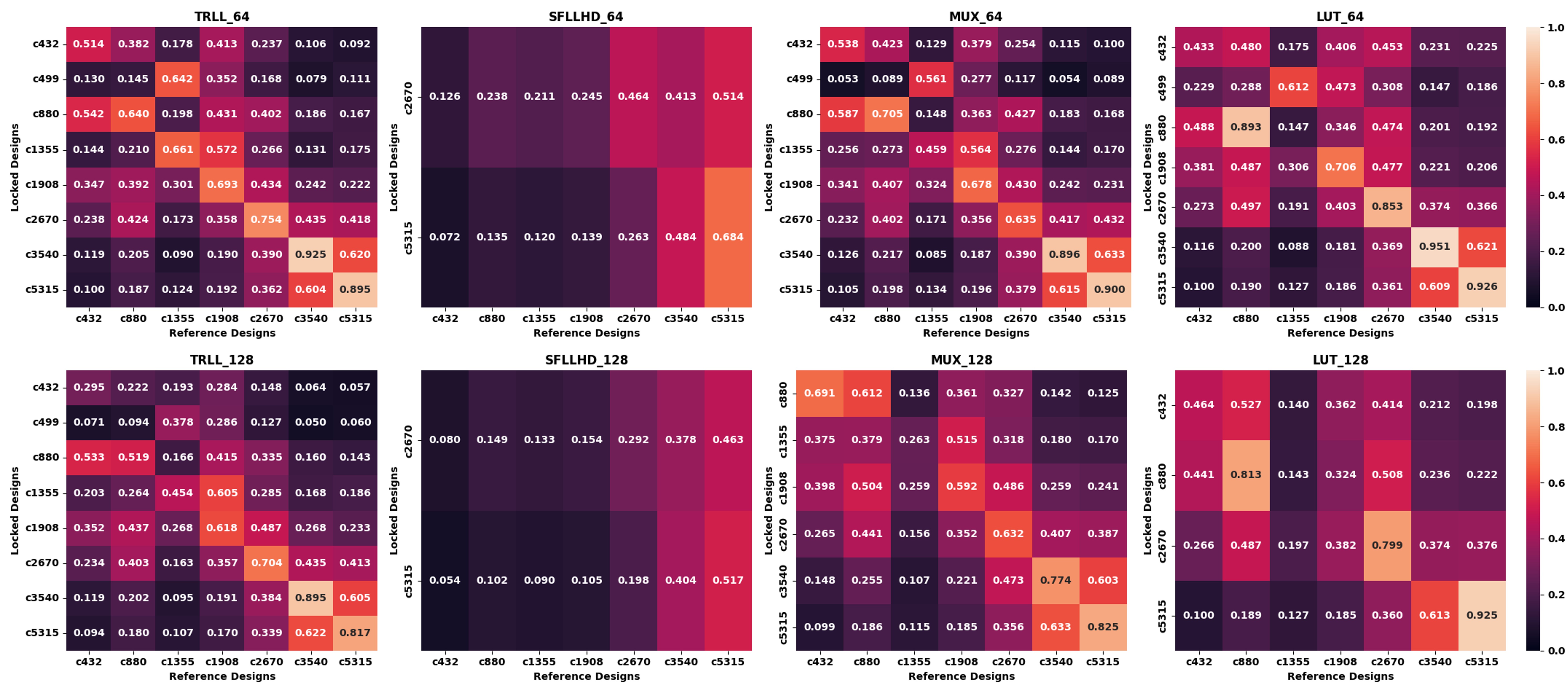}
        \caption{NPN similarity of 4-cut distributions between locked and reference designs. The top 20 cuts were taken for each gate. Top row: 64-bit keys, bottom row 128-bit keys.}
        \label{fig:heatmap_4cut}
        \vspace{-0.2in}
\end{figure*}

\begin{figure*}
        \centering
        \includegraphics[width=0.97\textwidth]{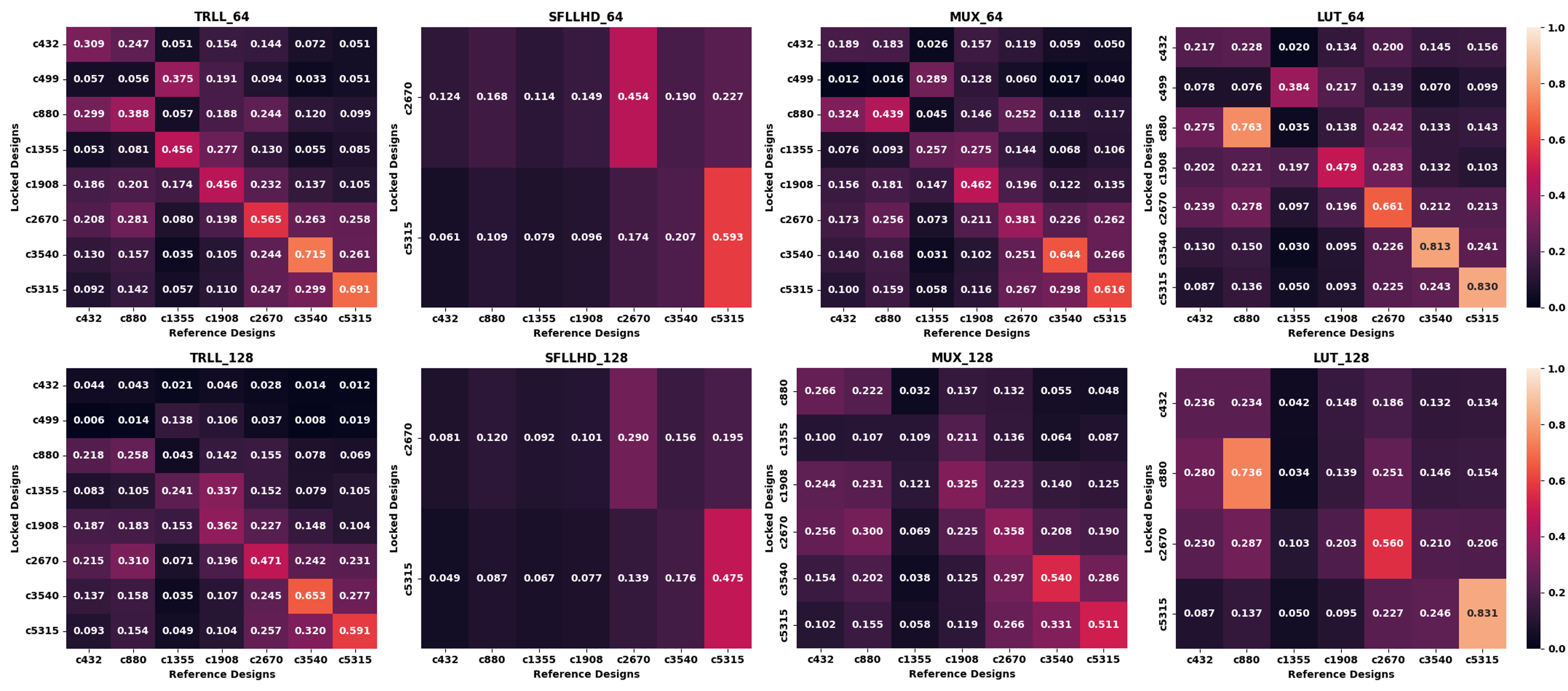}
        \caption{NPN similarity of 6-cut distributions between locked and reference designs. The top 20 cuts were taken for each gate. Top row: 64-bit keys, bottom row 128-bit keys.}
        \label{fig:heatmap_6cut}
        \vspace{-0.2in}
\end{figure*}

\begin{figure*}
        \centering
        \includegraphics[width=0.97\textwidth]{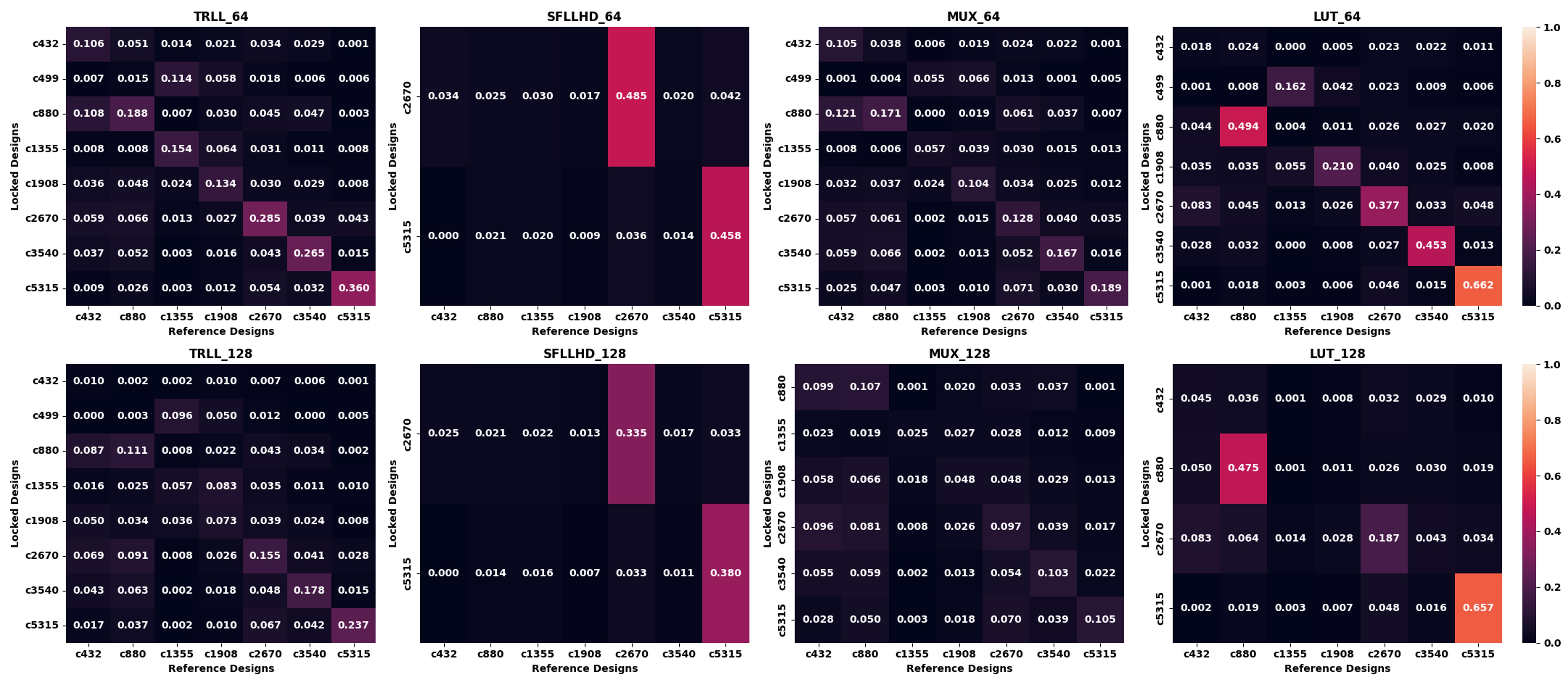}
        \caption{NPN similarity of 8-cut distributions between locked and reference designs. The top 20 cuts were taken for each gate. Top row: 64-bit keys, bottom row 128-bit keys.}
        \label{fig:heatmap_8cut}
        \vspace{-0.2in}
\end{figure*}

To further analyze the results, we plot heatmaps comparing the similarity scores for the locked benchmarks against all reference designs. Figures~\ref{fig:heatmap_4cut},~\ref{fig:heatmap_6cut},~\ref{fig:heatmap_8cut} illustrate the cut distribution similarity scores between locked and reference designs for top 20 cuts for 4-cut, 6-cut, and 8-cut distributions, respectively. Interestingly, while 4-cut has on average the highest matching similarity scores (73\%), it also has the overall lowest accuracy of the three cut sizes (84\%).  We attribute this observation to the fact that for 4-cuts, while there are $2^{2^4}$ possible functions, there are only 222 NPN classes~\cite{boolean2010}, which allows for higher overlap as evident in the similarity scores. 6- and 8-cut distributions do not have this problem. For example, the number of NPN equivalent classes for 6-cuts is $2.0\times10^{14}$~\cite{boolean2010}. This is evident in the lower similarities, but higher accuracies for the $k>4$ cut enumerations. Another observation is that for higher overhead locks, smaller benchmarks consistently struggle to be correctly identified. This is illustrated in Figures~\ref{fig:heatmap_4cut},~\ref{fig:heatmap_6cut},~\ref{fig:heatmap_8cut} by the diagonal not having the greatest similarity for smaller netlists. For example, with 128-bit MUX lock, the locked c880 benchmark is consistently incorrectly matched to c432, but with c880 a close second. Overall, while 8-cut provides the best accuracies on average (92\%), it has relatively low matching similarity scores (29\%), which calls its usefulness into question. 6-cut enumeration performs nearly as well (90\%) but with reasonably high matching similarity scores (54\%). 

\section{Discussion \& Limitations}
\label{sec:discussion}
While we show up to 92\% average accuracy in our results, we acknowledge several limitations of the proposed approach.
The current implementation is a coarse grained capture of $k$-cuts and is dependent upon the underlying structure in so far as it affects cut enumeration. The same design mapped to two different libraries can result in drastically different gate counts and connections. This is why we include the pre-processing step of resynthesizing both the L-DUT and reference designs to the same standard cell library using the same synthesis rules and prioritize selecting cuts with larger gate counts. Future work will look to further refine the cut enumeration process.
\begin{figure*}
        \centering
        \includegraphics[width=0.97\textwidth]{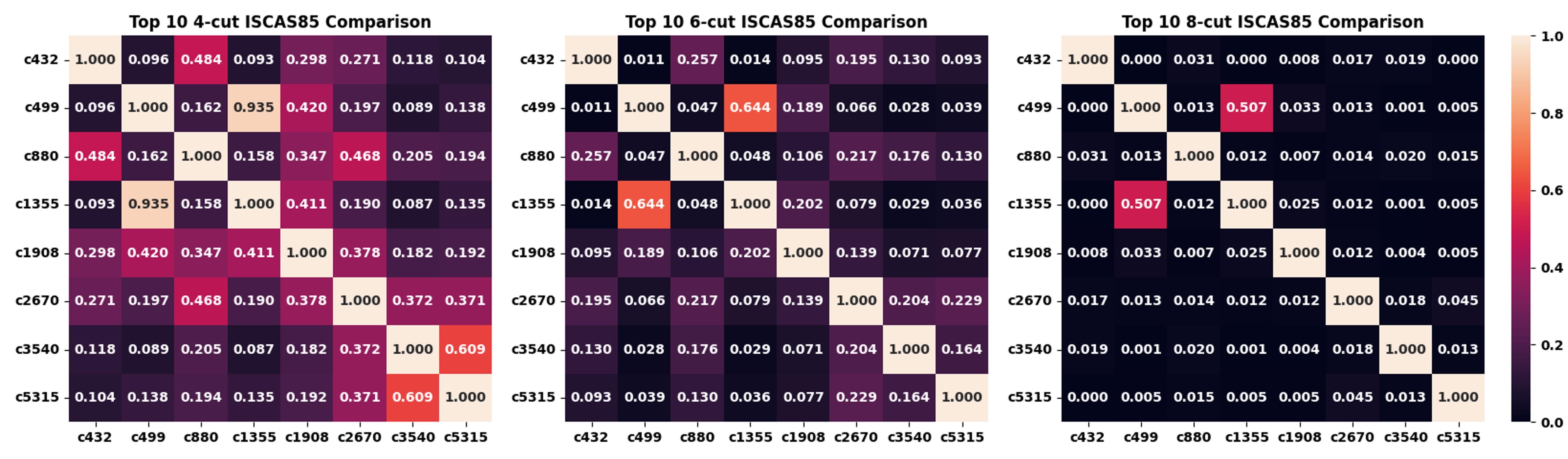}
        \caption{NPN similarity of 4-, 6- and 8-cut distributions between synthesized, unlocked ISCAS'85 benchmarks. The top 10 cuts were taken for each gate.}
        \label{fig:iscas_heatmap}
        \vspace{-0.25in}
\end{figure*}

The benchmark suite we tested is not very functionally diverse, with many designs sharing common functionality. For example, c499 and c1355 are functionally equivalent. c880, c2670, c3540, and c5315 are all described as ALUs with some commonalities among them~\cite{hansen2002unveiling}. We show the similarity heat maps among the benchmarks to highlight this in Figure~\ref{fig:iscas_heatmap}. Future work will look to evaluate our approach on more diverse and distinct benchmarks.

A final limitation involves interpreting the similarity score. Our evaluation determines correctness simply using the largest similarity value. However, in a non-test environment there is still the question of deciding whether a benchmark is deemed a match or if the original design does not exist in the current library. This problem is exacerbated in that for larger $k$-cuts, we observed low similarity scores. Likely, this value needs to be determined experimentally or supplemented with additional analysis. The authors in LATTE suggest performing other attacks if a likely benchmark is identified from chosen plaintext type analysis~\cite{dasgupta2025latte}.

\section{Conclusion} \label{sec:conclusion}
We proposed an approach for evaluating the indistinguishability of a logic lock using $k$-cut enumeration and Boolean NPN matching. Our approach is the newest in the line of database or chosen plaintext style attacks evaluating both the structural and functional transformations resulting from logic locking. We show the success of our technique through experimentation on several different types of locking, achieving a best case average accuracy of 92\%, even when the locked designs are resynthesized.

\section*{Acknowledgment}
\footnotesize
Sandia National Laboratories is a multi-mission laboratory managed and operated by National Technology \& Engineering Solutions of Sandia, LLC (NTESS), a wholly owned subsidiary of Honeywell International Inc., for the U.S. Department of Energy’s National Nuclear Security Administration (DOE/NNSA) under contract DE-NA0003525. This written work is authored by an employee of NTESS. The employee, not NTESS, owns the right, title and interest in and to the written work and is responsible for its contents. Any subjective views or opinions that might be expressed in the written work do not necessarily represent the views of the U.S. Government. The publisher acknowledges that the U.S. Government retains a non-exclusive, paid-up, irrevocable, world-wide license to publish or reproduce the published form of this written work or allow others to do so, for U.S. Government purposes. The DOE will provide public access to results of federally sponsored research in accordance with the DOE Public Access Plan.
\vspace{-0.1in}

\bibliographystyle{IEEEtran}
\bibliography{ref}
\end{document}